\documentclass[aps,floatfix,twocolumn]{revtex4}

\usepackage{amssymb}
\usepackage{graphics}
\usepackage{epsfig}
\usepackage{pgf,pgfarrows,pgfnodes,pgfautomata,pgfheaps}
\begin{document}

\title{Renormalization of tensor-network states}

\author{H. H. Zhao$^1$}
\author{Z. Y. Xie$^2$}
\author{Q. N. Chen$^2$}
\author{Z. C. Wei$^1$}
\author{J. W. Cai$^1$}
\author{T. Xiang$^{1,2}$}\email{txiang@aphy.iphy.ac.cn}

\affiliation{$^1$Institute of Physics, Chinese Academy of Sciences,
P.O. Box 603, Beijing 100190, China}

\affiliation{$^2$Institute of Theoretical Physics, Chinese Academy
of Sciences, P.O. Box 2735, Beijing 100190, China}

\date{\today}

\begin{abstract}
We have discussed the tensor-network representation of classical statistical or interacting quantum lattice models, and given a comprehensive introduction to the numerical methods we recently proposed for studying the tensor-network states/models in two dimensions. A second renormalization scheme is introduced to take into account the environment contribution in the calculation of the partition function of classical tensor network models or the expectation values of quantum tensor network states. It improves significantly the accuracy of the coarse grained tensor renormalization group method. In the study of the quantum tensor-network states, we point out that the renormalization effect of the environment can be efficiently and accurately described by the bond vector. This, combined with the imaginary time evolution of the wavefunction, provides an accurate projection method to determine the tensor-network wavfunction. It reduces significantly the truncation error and enables a tensor-network state with a large bond dimension, which is difficult to be accessed by other methods, to be accurately determined.
\end{abstract}


\maketitle

\section{Introduction}

The simulation of strongly correlated quantum or statistical
systems in two or higher dimensions remains a great challenge in physics. This has stimulated great interest on the investigation of so called tensor network states or models in recent years.\cite{Niggemann96, Verstraete04, Levin07, Jiang08, Xie09, Gu08, Hieida99} A tensor network state is a high-dimensional generalization of the one-dimensional matrix-product state\cite{Fannes92, Ostlund95} studied by the density matrix renormalization group (DMRG).\cite{White92} It captures accurately the key feature of entanglement in interacting quantum systems and is believed to be a good starting point for studying correlated systems. On the other hand, in a classical statistical system with local interactions, the Boltzmann weight can be expressed as a tensor product and all thermodynamic quantities can be determined by studying this equivalent tensor network model.

Strongly correlated quantum spin or fermions systems pose some of the most intriguing problems in condensed matter theory. Theoretical investigation on these problems generally starts from some simplified quantum lattice models, such as the Heisenberg model, which describes the interactions among local spins, or the Hubbard model, which describes the dynamics of band electrons subject to local Coulomb interactions. It is difficult to solve these models because the dimension of Hilbert space grows exponentially with the system size. Furthermore, as quantum fluctuations are strong in these systems, there are no obvious small parameters that can be used in perturbative expansions. This is the main obstacle to the resolution of many fundamental problems, such as high-T$_c$ superconductivity.

The quantum Monte Carlo and the DMRG are the two most commonly used numerical methods in the study of interacting quantum lattice models. The quantum Monte Carlo provides an accurate numerical tool for studying quantum spin models without frustrations, interacting boson models, or some special interacting fermion models. However, in most of fermion systems, such as the Hubbard model away from  half-filling, or quantum spin models with frustrations, such as the Heisenberg model on the Kagome lattice, the quantum Monte-Carlo methods are hampered by the ``minus-sign'' problem.

The DMRG was proposed by S. White to study ground states properties in 1992.\cite{White92} It has been then extended to study thermodynamic\cite{Nishino95, Bursill96, Wang97, Xiang99} as well as dynamic\cite{Hallberg95, Kuhner99, Luo03, Vidal04, White04, Daley04, Verstraete04PRL} properties of quantum lattice models. It can be also used to study a system with nonlocal interactions, either in the momentum space\cite{Xiang96} or other basis space\cite{White99}. An extensive review of DMRG is provided in Ref.~\cite{Schollwock05}. The DMRG minimizes the error in the truncation of basis states. It yields extraordinary precise results in one dimension. In two dimensions, the DMRG gives reasonably good results for small quantum systems.\cite{Liang94, Xiang01} However, its truncation error increases dramatically and the result becomes less reliable as the lattice size gets larger.

The DMRG is an iteration method. It starts by dividing a system, called a superblock, into two blocks. Like an experimental measurement, it uses one of the blocks as a needle to probe the states in the other block. The response function measured by the DMRG is the reduced density matrix. The DMRG works well in one dimension because it captures the main feature of the entanglement between the two blocks. The entanglement spectra is determined by the eigenvalues of the reduced density matrix, $\lambda_l$. The corresponding entanglement entropy is defined by
\begin{equation}
S = -\sum_l \lambda_l  \ln \lambda_l .
\end{equation}
As pointed out in Ref.~\cite{Xiang01}, the DMRG is to maximize the entanglement entropy during the basis truncation and can be regarded as a maximal entropy method.

For a gapped quantum system, the entanglement entropy is believed to satisfy an area law, namely the entanglement entropy scales with the cross section between the system and environment blocks\cite{Plenio05}. For a matrix product state, the entanglement entropy is bounded by
\begin{equation}
S \le \log D ,
\end{equation}
where $D$ is the matrix dimension. Thus to accurately represent a quantum state by a matrix product, the minimal dimension of the matrix, above which the wavefunction converges exponentially, must grow exponentially with the surface cross section. In one dimension, the surface contains just two points, a small number of states $D$ is sufficient to make DMRG extremely accurate.

In two or higher dimensions, the bond dimension that is needed for describing accurately a ground state grows exponentially with the system size. It is difficult to use a matrix product wavefunction to represent faithfully a quantum state in two or higher dimensions. That is why the DMRG can yield reliable results only in small lattice systems in two dimensions. It is an intrinsic barrier in the application of the DMRG in two or higher dimensions.

At a critical point, there is a logarithmic correction to the area law.\cite{Vidal-area03,Calabrese04} This logarithmic correction does not affect much on the calculation of short range correlation functions. However, as it diverges with the system size, it can affect strongly on the calculation of long range correlation functions. The minimal matrix dimension will grow exponentially with the system size. This is why the DMRG works better in a gapped system, for example the spin-1 Heisenberg model, than in a critical system, for example the spin-1/2 Heisenberg model.

In 1995, Ostlund and Rommer pointed out that the wavefunction generated by the DMRG iteration is a matrix-product state.\cite{Ostlund95} A matrix product state can be also taken as a variational wavefunction. It can be accurately determined by the DMRG in one dimension.

The DMRG is a one-dimensional algorithm. To apply the DMRG in more than one dimension, one needs to map a two- or higher-dimensional lattice into a one-dimensional one by introducing nonlocal coupling.\cite{Xiang01} However, locality is the key of the great performance of the DMRG in one dimension. To capture the entanglement between any two neighboring sites, one needs to keep the locality of interactions.

A tensor-network state is a high-dimensional extension of the one-dimensional matrix-product state. It keeps the locality of interactions and captures the main feature of the area law\cite{Verstraete04}, since the number of entangled bonds between the systems and environment blocks is proportional to the interface area. A tensor product representation reflects accurately the short-range entanglement of a quantum state. Like in a matrix product state in one dimension, the minimal bond dimension that is needed for accurately describing a quantum  state does not depend on the system size in a gapped system. This is an advantage of using the tensor representation.

The tensor-network representation of quantum states is in fact not new. A typical example of tensor-network state is the valence-bond-solid (VBS) state that was proposed by Afflect et. al more than two decades ago\cite{AKLT87}. It is the exact ground state of a class of two-dimensional quantum antiferromagnets. A tensor product wavefunction was also constructed by Niggemann et al. to study the ground state properties of the Heisenberg model on the honeycomb lattice\cite{Niggemann96}. They pointed out that the calculation of expectation values of tensor product states is equivalent to evaluating a classical partition function. A more general ansatz of tensor network states was suggested by Sierra and Martin-Delgado\cite{Sierra98}. The use of the tensor network state as a variational wavefunction for the three-dimensional classical lattice model was suggested by Nishino.\cite{Nishino01} In 2004, the idea of tensor network state was discussed by Cirac and coworkers under the name of projected entangled-pair states (PEPS)\cite{Verstraete04}. Their work has attracted great attention because it reveals more clearly the physical picture embedded in the tensor-network representation of quantum states.

Even before the tensor-network wavefunction was introduced in quantum systems, a tensor-network representation of the partition function for classical lattice models has been widely used in statistical physics. A vertex model\cite{Baxter82} is such an example. Furthermore, as will be discussed in Sec.~\ref{sec:TPS}, all classical lattice models with local interactions can be written as tensor-network models. Unlike in a quantum system, the local tensor can be readily determined from its Hamiltonian.

In the past decade, a number of variational approaches have been suggested to determine the tensor-product wavefunction in quantum systems.\cite{Murg07, Sandvik07} However, the number of free tensor elements that can be handled by a variational approach is too small to resolve any physical problems that are difficult to be resolved by other methods. This is because in a variational approach, the maximal number of free variables can be accurately and efficiently determined from the minimization of the expectation value of Hamiltonian is generally less than $10^2\sim 10^3$. This has limited the bond dimension to be generally less than 5.

In Ref.~\cite{Jiang08}, we proposed a projection method (or power method) to evaluate the tensor-network wavefunction of the ground state. In this method, the projection operator is applied iteratively to a random initial wavefunction by utilizing the Trotter-Suzuki decomposition. The contribution of the environment is approximately taken into account by the diagonal bond matrix.  This method is quite efficient and accurate, since in the evaluation of the wavefunction by the imaginary time evolution, the Trotter and truncation errors do not accumulate in the iteration. At each step of iteration, a singular value decomposition for a matrix of dimension $d D^2$ on a honeycomb lattice, or $d D^3$ on a square or Kogome lattice, needs to be done. Here $d$ is the dimension of the local basis set and $D$ is the bond dimension of the local tensor. For the spin-1/2 Heisenberg model $d = 2$. Nowadays a matrix of dimension $\sim 10^4$ can be efficiently diagonalized by a desktop computer. Therefore, a tensor-network state with the bond dimension up to $D=70$ on a honeycomb lattice, or $D=17$ on a square or Kagome lattice, can be calculated for the S=1/2 Heisenberg model.

The above projection method is similar that used by Vidal in one dimension\cite{Vidal07}. But there is a significant difference between one and two dimensions. In one dimension, there is a canonical representation of the matrix product state in which the bond vector is preciously the singular spectrum of Schmidt decomposition. If the matrix product state is canonical, then by cutting a bond, the left and right matrix product states are strictly orthogonal. The projection method introduced in Ref.~\onlinecite{Vidal07} can be naturally imposed by applying the orthogonality condition. However, in two dimensions, this kind of orthogonality condition does not exist and the bond vector is no longer the singular spectrum of any kind of Schmidt decompostion since to separate a point one needs to cut at least four bonds on a square lattice. Our approach is to take the bond vector as an effective entanglement measure of environment. It goes beyond the canonical representation of tensor-network wavefunctions.

To evaluate the expectation values of physical observables from a given tensor-network wavefunction, a number of approaches, including  the transfer matrix renormalization group introduced in Refs.~\cite{Bursill96, Wang97, Xiang99} or the variational Monte Carlo mehtods\cite{Sandvik07}, can be used. In Ref.~\cite{Jiang08}, we adopted the coarse graining tensor renormalization group (TRG) method proposed by Levin and Nave \cite{Levin07}. From the calculation, we find that the TRG can indeed produce qualitatively correct results when the bond dimension of tensors is small. However, the truncation error in the TRG iteration grows rapidly with the bond dimension of local tensors. This leads to a big error in the calculation of expectation values. In particular, the ground state energy and other physical quantities do not converge in the large bond dimension limit.

In the TRG method of Levin and Nave, the singular-value spectra of a local matrix $M$ defined by a product of two neighboring local tensors is renormalized in the truncation of basis space. This provides a local optimization of the truncation space since it does not consider the influence of the environment tensors which are defined from the whole lattice by excluding the two tensors on which $M$ is defined. Similar as in the DMRG, the interplay between the system ($M$ here) and environment tensors can modify dramatically the singular-value spectra of $M$. Thus the renormalization effect of the environment to $M$ should be considered to globally optimize the accuracy of physical observable.

In Ref.~\cite{Xie09}, we introduced a second renormalization method of tensor network model (abbreviated as SRG) to study this renormalization effect of environment. This method can be used to evaluate the partition functions of classical statistical models and the expectation values of quantum tensor-network wavefunctions. It improves significantly the accuracy of TRG. For example, for the two-dimensional Ising model, we found that this method can improve the accuracy of the TRG by more than two orders of magnitude in the vicinity of the critical point and more than five orders of magnitude away from the critical point by keeping $D_{cut}= 24$ states. This method also provides a powerful numerical tool for accurately evaluating the expectation values of tensor-network states in a quantum lattice model. In contrast to the TRG, we find that the SRG results for the ground state energy of the Heisenberg model on a honeycomb lattice converge quickly with increasing bond dimension $D$.

The SRG together with the projection method introduced in Ref.~\cite{Jiang08} establishes an efficient numerical technique for studying the tensor network states of quantum lattice models in two dimensions. The SRG alone also provides an accurate numerical tool for studying thermodynamics of classical lattice models. In this paper, we will give a comprehensive introduction to these methods.

This paper is arranged as follows. Sec.~\ref{sec:TPS} is devoted to a discussion on the tensor-network representation of classical lattice models. Two methods are introduced to convert a classical lattice model into a tensor-network model. One is to define the local tensor in the dual lattice by performing a duality transformation, and the other is to define the local tensor by singular value decomposition in the original lattice. Sec.~\ref{sec:SRG} starts by a brief review of the TRG method. A detailed introduction to the SRG is then given. Sec.~\ref{sec:quantum} describes the iterative projection method for determining the tensor-network wavefunction of ground state in a quantum system. The calculation of the expectation values of tensor-network states is discussed in Sec.~\ref{sec:expect}. Sec.~\ref{sec:summary} is a summary.

For reference, we summarize a few notations used in this work: $d$ is the dimension of local physical basis set. $D$ is the bond dimension of a matrix or tensor product state in a quantum system. $D_c$ is the bond dimension of a classical tensor network model. In a quantum system, $D_c=D^2$ is the bond dimension of the tensor used in the evaluation of physical expectation values. $D_{cut}$ is the bond dimension of the tensor retained in the TRG or SRG calculations.

\section{Tensor network representation of classical statistical models }
\label{sec:TPS}

In this section, we discuss about the tensor-network representation of classical statistical models with local interactions. It is well known that the one-dimensional spin-1/2 Ising model can be rigorously solved by expressing its partition function as a product of transfer matrix. This is the simplest example of tensor network representation of classical statistical model. A matrix is a order two tensor. A tensor network representation of classical statistic model is a higher dimensional extension of the one-dimensional matrix product. All statistical models with short-range interactions, such as the Ising model and the Pott's model, can be expressed as tensor network models.

There are two approaches to represent the partition function of a classical statistical model in a tensor-network form. One is to take a duality transformation of the model and define the tensor in the dual space.\cite{Levin07} This can transform a classical statistical model to a tensor-network model in its corresponding dual lattice. The order of the local tensor is the coordinate number of the dual lattice. This kind of transformation is particularly useful if the coordinate number of the dual lattice is smaller than the coordinate number of the original lattice. For example, a classical model defined on a triangular lattice whose coordinate number is 6 can be represented as a tensor-network model on a honeycomb lattice whose coordinate number is 3. The dual lattice of a dice lattice is a Kagome lattice. The square lattice is self-dual.

If the original statistical model includes not only the nearest neighboring two-site interactions, but also many-site interactions within each constitutional unit cell, for example three- or four-site interaction within each plaquette in a square lattice, a tensor-network model can still be defined in the dual lattice without enlarging the bond dimension of tensor. This is another advantage for representing the partition function in the dual space.

The second approach is to define the tensor in the original lattice by taking singular value decompositions for the local bond partitions. This approach can be also implemented in all kinds of lattices. The order of the tensor is equal to the coordinate number of the lattice if there are only interactions between neighboring sites. This kind of representation is more appropriate for a lattice, such as a honeycomb lattice, whose coordinate number is small.

\subsection{Tensor network representation in the dual lattice}

Let us take the $S=1/2$ Ising model defined on a triangular lattice as an example to illustrate how to express its partition function as a tensor product in its dual lattice. The Hamiltonian of the Ising model is defined by
\begin{equation}
H = -J \sum_{\langle ij \rangle } S_i S_j,
\end{equation}
where $S_i$ takes two values $\pm 1$. This model is rigorously soluble\cite{Wannier}.

The partition function of the Ising model is given by
\begin{eqnarray}
Z & = & \mathrm{Tr} e^{-\beta H}  \nonumber \\
& = & \mathrm{Tr} \prod_{\triangle_{ijk}} e^{\beta J (S_i S_j +
S_j S_k + S_k S_i) / 2 ,}
\end{eqnarray}
where $\mathrm{Tr}$ is to sum over all spin configurations and the product is taken over all small triangles. The factor $1/2$ is introduced in the above exponent because each bond is shared by two triangles. This partition function can be expressed as a tensor network in terms of bond variables through a duality transformation. The dual lattice of a triangular lattice is a honeycomb lattice.

Given a bond on the triangular lattice, let us define the bond spin by the the product of two end spins as
\begin{equation}
\label{eq:sigma}
\sigma _{ij}=S_{i}S_{j}.
\end{equation}
This bond spin takes two values $\sigma =1$ or $-1$, corresponding to a state of parallel or antiparallel Ising spins. The partition function can then be expressed as
\begin{equation}
Z=\mathrm{Tr}\prod_{\langle ij\rangle
}\delta (\sigma _{ij}-S_{i}S_{j})\prod_{\triangle _{ijk}}e^{\beta
J(\sigma _{ij}+\sigma _{jk}+\sigma _{ki})/2},
\end{equation}
where $\mathrm{Tr}$ is to sum over all $S$ and $\sigma$ spins.

Since the number of the bonds ($N_{bond}$) is equal to the sum of the number of sites ($N_{site}$) and the number of triangles ($N_{triangle}$), the $ N_{bond}$ delta functions in the above equation can then be equivalently written as a product of $N_{site}$ delta functions defined on all lattice sites and $N_{triangle}$ delta functions defined on all triangles. On each triangle, it is simple to show that the product of the three bonds spins is equal to one
\begin{equation}
\sigma _{ij}\sigma _{jk} \sigma_{ki} = S_{i}
S_{j} S_{j} S_{k} S_{k} S_{i} = 1.
\end{equation}
This is the constraint defined on each triangle, independent on the original spin variables. Thus the $N_{site}$-site delta functions can be integrated out. The partition function then becomes
\begin{equation}
Z=\mathrm{Tr}_{\sigma }\prod_{\triangle _{ijk}}\frac{1+\sigma
_{ij}\sigma _{jk}\sigma _{ki}}{2}e^{\beta J(\sigma _{ij} + \sigma _{jk}+\sigma _{ki})/2}.
\label{eq:ising1}
\end{equation}

Now if use $i$ to denote the site position of the dual lattice, then the above partition function can be expressed as a standard tensor-network model
\begin{equation}
Z=\mathrm{Tr} \prod_{i}T_{x_{i}y_{i}z_{i}},
\label{eq:tnIsing}
\end{equation}
where the trace is to sum over all indices and
\begin{equation}
T_{x_{i}y_{i}z_{i}}=\frac{1+\sigma _{x_{i}}\sigma _{y_{i}}\sigma_{z_{i}}}{2} e^{\beta J(\sigma _{x_{i}}+\sigma _{y_{i}}+\sigma _{z_{i}})/2}
\end{equation}
is a third-order tensor defined on the hexagonal lattice. $x_{i}$, $y_{i}$, and $z_{i}$ are the three integer bond indices of dimension $D_c = 2$ defined on the three bonds emitted from site $i$ along the $x$, $y$, and $z$ directions, respectively. Each bond links two sites. The two bond indices defined from the two end points take the same values.

The above derivation can be extended to other lattices. A square lattice is self-dual. For the spin-1/2 Ising model, it is straightforword to show that the tensor-network representation of the partition function on the square is given by
\begin{equation}
Z=\mathrm{Tr}_{\sigma }\prod_{i} T_{x_{i}y_{i} x^{\prime
}{}_{i}y_{i}^{\prime }},
\label{eq:square}
\end{equation}
where
\begin{equation}
T_{x_{i}y_{i}x^{\prime }{}_{i}y_{i}^{\prime }}=\frac{1+\sigma
_{x_{i}}\sigma _{y_{i}}\sigma _{x^{\prime }{}_{i}}\sigma
_{y_{i}^{\prime }}}{2}e^{\beta J(\sigma _{x_{i}}+\sigma
_{y_{i}}+\sigma _{x^{\prime }{}_{i}}+\sigma _{y_{i}^{\prime }})/2} .
\end{equation}
is a fourth-order tensor defined on the dual square lattice. $x_{i}$, $y_{i}$ , $x^{\prime }{}_{i}$ and $y_{i}^{\prime }$ are the four bonds connecting site $i$.

For the honeycomb lattice, its dual lattice is a triangular lattice. In the dual space, each site has six neighbors. The order of the local tensor is six. Thus there is no advantage to carry out the calculation in the dual space.

\begin{figure}
\includegraphics[width=0.45\textwidth]{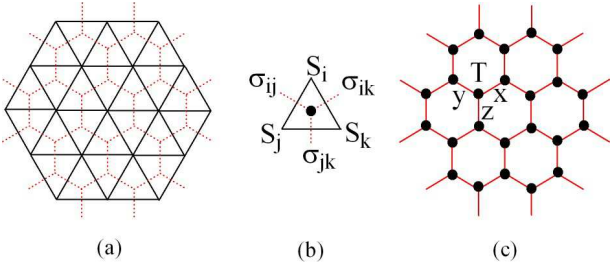}
\caption{\label{lattice} (color online) Schematic representation of the partition function of a classical system defined by Eq.~(\ref{eq:ising1}) on the dual lattice of the triangular lattice, namely the honeycomb lattices. }
\end{figure}

The above duality transformation can be extended to other statistical models. For example, the $q$-state Pott's model is defined by the Hamiltonian on a triangular lattice
\begin{equation}
H = J \sum_{\langle i,j\rangle} \delta (\theta_i - \theta_j),
\end{equation}
where $ \theta_i = 0, 1, \cdots q-1$ and $\delta (a-b)$ is the Kronecker delta function, it is straightforward to show that its partition function can be also written as a tensor product in the dual lattice. But the local tensor is now defined by
\begin{equation}
T_{xyz}=\delta (\textrm{mod}[x+y+z, q]) e^{-\beta J [\delta (x) +
\delta (y) + \delta (z)]/2},
\end{equation}
where the bond variable $x$ (similarly $y$ or $z$) takes integer values from 0 to $q-1$. When $q=2$, the Potts model is equivalent to the spin-1/2 Ising model.

However, it should be pointed out that Eq.~(\ref{eq:sigma}) is not a one-to-one mapping. The above derivation can be applied to the Ising model without magnetic field. If a Zeeman term or a many-site interaction term is added to the Hamiltonian, then the transformation Eq.~(\ref{eq:sigma}) can no longer be used. Nevertheless, the tensor-network model can still be defined in the dual lattice. But the dimension of the tensor has to be extended from $D_c=d$ ($d=2$ for the spin-1/2 Ising model) to $D_c= d^2=4$ to distinguish the 4 spin configurations of $S_i$ and $S_j$.

\begin{figure}[tb]
\includegraphics[width=0.15\textwidth]{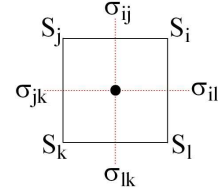}
\caption{(color online) A unit cell of a square lattice and its corresponding dual variables.}
\label{fig:square}
\end{figure}

To understand this more concretely, let us consider an extended spin-1/2 Ising model in an applied magnetic field defined on a square lattice
\begin{equation}
H = -J \sum_{\langle ij\rangle} S_iS_j - h \sum_i S_i + J_{\square} \sum_{ijkl \in \square} S_iS_jS_k S_l,
\end{equation}
where the last term is to sum over all four-spin interaction terms defined on each plaquette. $ijkl \in \square$ means that $ijkl$ are the four-vertex indices on each plaquette, as shown in Fig.~\ref{fig:square}. The partition function of this model can still be expressed as a tensor product defined by Eq.~(\ref{eq:square}). But the local tensor $T_{x_iy_ix_i^\prime y_j^\prime}$ is now defined by
\begin{eqnarray}
T_{\sigma_{il} \sigma_{ij} \sigma_{jk} \sigma_{lk}}
& = &\exp (-\beta H_{\square}) \delta_{\sigma_{il}, I_{il}} \delta_{\sigma_{ij}, I_{ij}}
\nonumber \\
&& \delta_{\sigma_{jk}, I_{jk}} \delta_{\sigma_{lk}, I_{lk}} ,
\end{eqnarray}
where
\begin{eqnarray}
H_\square &=& -\frac{J}{2} \left(S_iS_l + S_iS_j + S_jS_k + S_lS_k\right)
\nonumber \\
&&
- \frac{h}{4} \left( S_i + S_j + S_l + S_k\right) + J_{\square} S_iS_jS_k S_l.
\end{eqnarray}
If we assume $S_i=\pm 1$ for the spin-1/2 Ising model, then $I_{ij}$ is defined by
\[
I_{ij}  =  (S_i+1) + (S_j+1)/2 .
\]
It takes four integer numbers from 0 to 3. The bond dimension of the tensor is $D_c=4$.

\subsection{Tensor network representation in the original lattice}

A classical statistical model with local interactions in any spatial dimension can be always expressed as a tensor-network model in its original lattice. To do this, let us consider a model defined by the following Hamiltonian
\begin{equation}
H = \sum_{\langle ij \rangle} K(S_i, S_j),
\end{equation}
where $\langle\,\, \rangle$ denotes summation over nearest neighbors only. $K(S_i,S_j)$ is the interaction between two local basis states $S_i$ and $S_j$. The partition function of this model is given by
\begin{equation}
Z = \sum_{\{S_i\}} \prod_{\langle ij \rangle} W(S_i, S_j) ,
\end{equation}
where
\begin{equation}
W(S_i, S_j) = \exp \left[ -\beta K(S_i, S_j) \right]
\end{equation}
defines a matrix whose row and column indices are $S_i$ and $S_j$, respectively. $W$ is not necessary to be symmetric and positive defined.

Given $W(S_i,S_j)$, one can take a singular value decomposition to decouple it into the following form
\begin{equation}
W(S_i, S_j) = \sum_l U(S_i, l) \lambda_l V(S_j, l),
\end{equation}
where $U$ and $V$ are unitary matrices, $\lambda$ is a
semi-positive diagonal matrix. If we define
\begin{eqnarray}
Q_a (S,l) &= & U(S,l) \lambda^{1/2}_l,
\\
Q_b (S,l) &=& V(S, l) \lambda^{1/2}_l,
\end{eqnarray}
then $W$ can be reexpressed as
\begin{equation}
W = Q_a Q_b .
\end{equation}

Now let us group all $Q$'s that connect to site $i$. A local tensor can then be defined by tracing out $S_i$ from the product of these $Q$'s
\begin{equation}
T^i_{x,y,z,\dots} = \sum_{S_i} Q_{\alpha_1}(S_i,x)
Q_{\alpha_2}(S_i,y) Q_{\alpha_3}(S_i,z) \cdots .
\end{equation}
The order of the tensor is equal to the number of bonds interacting with site $i$. The bond dimension is equal to the site dimension $d$. This gives a tensor-network representation of the partition function
\begin{equation}
Z = \mathrm{Tr}\prod_{i} T^i_{x_{i},y_{i},z_{i}\dots} .
\end{equation}
The two bond indices defined from the two end points take the same
values. The trace is to sum over all bond indices.

On a bipartite lattice, for example a honeycomb lattice, the partition function can be simply expressed as
\begin{equation}
\label{eq:TaTb}
Z = \mathrm{Tr}\prod_{i\in a,j\in b} T_{x_{i},y_{i},z_{i}}^{a}
T_{x_{j},y_{j},z_{j}}^{b} ,
\end{equation}
where the superscripts $a$ and $b$ stand for the two sublattices of the honeycomb lattice, and
\begin{eqnarray}
T_{x,y,z}^{a} &=&\sum_{S}Q_{a}(S,x)Q_{a}(S,y)Q_{a}(S,z), \\
T_{x,y,z}^{b} &=&\sum_{S}Q_{b}(S,x)Q_{b}(S,y)Q_{b}(S,z).
\end{eqnarray}
Moreover, if $W$ is positive and symmetric, then $V=U$ and $Q_{a}=Q_{b}$. In this case $T\equiv T^{a}=T^{b}$ and the partition funciton can be simply expressed as
\begin{equation}
Z = \mathrm{Tr} \prod_{i} T_{x_{i},y_{i},z_{i}} .
\end{equation}

\begin{figure}
\includegraphics[width=0.45\textwidth]{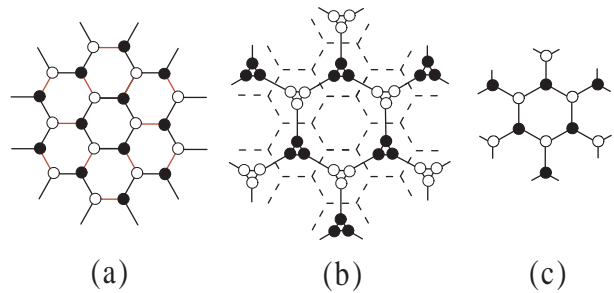}
\caption{Rewiring of a honeycomb lattice by the singular value decomposition in the TRG iteration. }
\label{fig:rewiring}
\end{figure}

\section{Second Renormalization of Tensor Network Model}
\label{sec:SRG}

\subsection{Overview of TRG}

We first review briefly the TRG method\cite{Levin07}. Let us take the model defined by Eq.~(\ref{eq:TaTb}) on a honeycomb lattice as an example to show how the method works.

\begin{figure}
\includegraphics[width=0.4\textwidth]{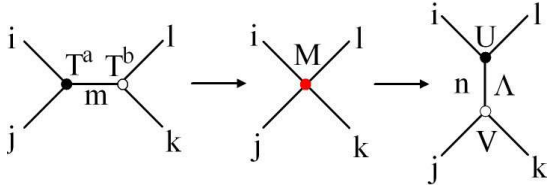}
\caption{Schematic representation of Eqs.~(\ref{eq:M}) and (\ref{eq:Msvd}). }
\label{fig:svdm}
\end{figure}

The TRG is an iterative method. There are two steps at each iteration. The first step is to rewire a honeycomb lattice to a triangle-honeycomb lattice as shown in Fig.~\ref{fig:rewiring}(b). This is done by transforming a pair of neighboring tensors, $T^a$ and $T^b$, into two new tensors $S^a$ and $S^b$ defined in the rewired lattice, by performing a singular value decomposition as schematically shown in Fig.~\ref{fig:svdm}. The common bond index of $T^a$ and $T^b$ is first contracted to form a matrix $M$ defined by
\begin{equation} \label{eq:M}
M_{li , jk} = \sum_{m} T^a_{mij} T^b_{m kl} .
\end{equation}
The dimension of $M$ is the product of the corresponding bond dimensions. The initial dimension of $M$ is $D_c^2$. The singular value decomposition is then applied to decouple this matrix into the following form
\begin{equation}
\label{eq:Msvd}
M_{li, jk} = \sum_{n= 1} U_{li , n} \Lambda_n V_{jk , n},
\end{equation}
where $U$ and $V$ are two unitary matrices. $\Lambda_n$ is a
semi-positive diagonal matrix arranged in descending order.
It measures the entanglement between $U_n$ and $V_n$. The dimension of $\Lambda$ is equal to that of $M$, higher than the original bond dimension. To carry out the calculation iteratively, one has to truncate the basis space to a manageable level and retain $D_{cut}$ largest singular values and the corresponding vectors. After that, one can define two new tensors
\begin{eqnarray}
S^{a}_{nli} &=& U_{li,n}\sqrt{\Lambda_n} ,
\label{eq:Sa} \\
S^{b}_{njk} & = & V_{jk,n}\sqrt{\Lambda_n}.
\label{eq:Sb}
\end{eqnarray}

\begin{figure}
\includegraphics[width=0.35\textwidth]{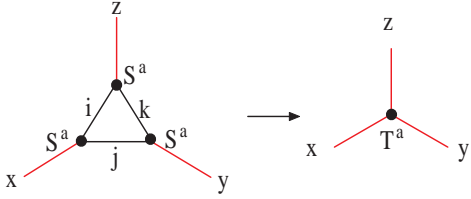}
\caption{Schematic representation of Eq.~(\ref{eq:decimation}). A local tensor $T^a$ is defined by contracting the three internal bonds on each triangle.}
\label{fig:decimation}
\end{figure}

The second step is to contract each small triangle in the rewiring
lattice to define two new local tensors in the squeezed honeycomb lattice (Fig.~\ref{fig:decimation})
\begin{equation} \label{eq:decimation}
T^\alpha_{xyz} = \sum_{ijk} S^\alpha_{xji} S^\alpha_{ykj} S^\alpha_{zik},
\end{equation}
where $\alpha = a$ or $b$. After that, the lattice size is reduced by a fact of 3.

\begin{figure}[b]
\includegraphics[width=0.2\textwidth]{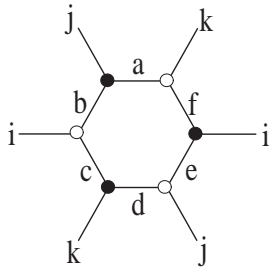}
\caption{The last six sites in the TRG iteration. A central symmetric boundary condition is assumed.  }
\label{fig:six}
\end{figure}

By repeating the above steps iteratively, one can finally reach a hexagonal lattice with only six tensors. The partition function can then be evaluated by tracing out all bond indices of these six tensors, assuming a central symmetric boundary (Fig.~\ref{fig:six}).

\subsection{SRG}

\begin{figure}
\includegraphics[width=0.4\textwidth]{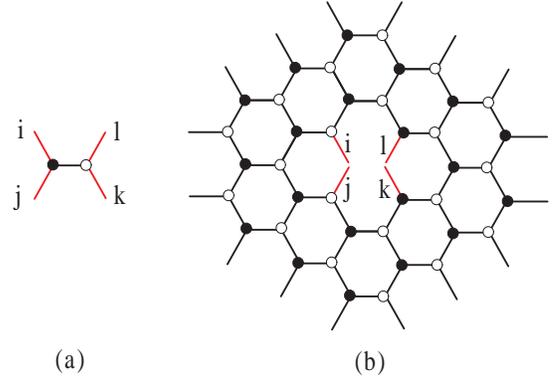}
\caption{Configuration of a system (a) and its corresponding environment lattice (b).}
\label{fig:Env0}
\end{figure}

The above TRG iteration minimizes the truncation error of local matrix $M$. However, as pointed out in Ref.~\cite{Xie09}, it does not consider the influence of the environment which contains all the lattice points excluding the two on which $M$ is defined (Fig.~\ref{fig:Env0}). In real calculation, it is the truncation error of the partition function rather than that of the local matrix $M$ that should be minimized. This means that the TRG is just a local optimization method.

To optimize the partition function globally, one needs to consider the renormalization effect of the environment to $M$. In Ref.~\cite{Xie09}, this is called the second renormalization method of tensor network model (SRG). This SRG method, as demonstrated in Ref.~\cite{Xie09}, can incorporate efficiently the renormalization effect of environment and improve significantly the TRG method.

The difference between the TRG and SRG is similar to the difference between the conventional Wilson block renormalization group method\cite{Wilson74} and the DMRG\cite{White92}. In the conventional block renormalization group method, it is the block Hamiltonian that is optimized without considering the interaction between different blocks. However, in the DMRG, the basis states of the system block are optimized by fully considering the interplay between the system and environment blocks via the reduced density matrix. The singular values in the DMRG are the coefficients of the Schmidt decomposition of the density matrix. They measure the entanglement between system and environment blocks.

This can be understood more clearly by expressing the partition function as a product of $M$ and its corresponding environment matrix $M^e$
\begin{equation}
Z = TrMM^{e} ,
\end{equation}
where $M^{e}$ is defined by contracting over all bond indices in the environment lattice. From this formula, it is clear that to reduce the error in $Z$, one needs to minimize the truncation error of $MM^{e}$, not just that of $M$.

We will discuss how to determine the values of $M^e$ later. Once $M^{e}$ is known, its renormalization to $M$ can be done in the following steps:

1. To take a singular value decomposition for $M^e$
\begin{equation}
M_{jk,li}^{e}={\sum_{n} }U_{jk,n}^{e}\Lambda _{n}^{e}V_{li,n}^{e},
\end{equation}
where $U^{e}$ and $V^{e}$ are two unitary matrices and $\Lambda^{e}$ is a semi-positive diagonal matrix. This step is taken to ensure that the renormalization effect of $M^e$ to $M$ can be more symmetrically treated and the truncation error of the partition function is minimized.

2. From the above decomposition, one can define a new matrix:
\begin{equation}
\label{eq:Me}
\tilde{M}_{n_{1},n_{2}}={\sum_{lijk} }\left( \Lambda
_{n_{1}}^{e}\right) ^{1/2}V_{li,n_{1}}^{e}M_{li,jk}U_{jk,n_{2}}^{e}\left(
\Lambda _{n_{2}}^{e}\right) ^{1/2},
\end{equation}
and represent the partition function as
\begin{equation}
Z = Tr\tilde{M}.
\end{equation}
This equation means that the error of the partition is minimized if the truncation error of $\tilde{M}$ is minimized. Now we perform a singular value decomposition for $\tilde{M}$
\begin{equation}
\label{eq:Mtilde-svd}
\tilde{M}_{n_{1},n_{2}}={\sum_{n} }\tilde{U}_{n_{1},n}\tilde{\Lambda}_{n}\tilde{V}_{n_{2},n},
\end{equation}
again. $\tilde{U}$ and $\tilde{V}$ are two unitary matrices, and $\tilde{\Lambda}$ is a semi-positive diagonal matrix. According to the least square principle, the truncation error of $\tilde{M}$ is minimized if the $D_{cut}$ largest singular values of $\tilde{\Lambda}$ are retained in the truncation.

3. By substituting the truncated $\tilde{M}$ back to Eq.~\ref{eq:Me}, one can represent $M$ as
\begin{equation}
M_{li,jk}={\sum_{n_{1}n_{2}} }V_{li,n_{1}}^{e}\left( \Lambda
_{n_{1}}^{e}\right) ^{-1/2}\tilde{M}_{n_{1},n_{2}}\left( \Lambda
_{n_{2}}^{e}\right) ^{-1/2}U_{jk,n_{2}}^{e}.
\end{equation}
It can be further expressed as a product of two tensors
\begin{equation}
M_{li,jk}\approx \sum_{n=1}^{D_{cut}}S_{n,li}^{a}S_{n,jk}^{b} ,
\end{equation}
where
\begin{eqnarray}
S_{n,li}^{a}={\sum_{n_{1}} }V_{li,n_{1}}^{e}\left( \Lambda
_{n_{1}}^{e}\right) ^{-1/2}\tilde{U}_{n_{1},n}\left( \tilde{\Lambda}%
_{n}\right) ^{1/2} ,\\
S_{n,jk}^{b}={\sum_{n_{2}} }U_{jk,n_{2}}^{e}\left( \Lambda
_{n_{2}}^{e}\right) ^{-1/2}\tilde{V}_{n_{2},n}\left( \tilde{\Lambda}%
_{n}\right) ^{1/2},
\end{eqnarray}
are the two tensors defined in the rewired lattice.

Then one can follow the steps of TRG to update tensors $T^a$ and $T^b$ in the squeezed lattice by taking the coarse grain decimation of $S^a$ and $S^b$. This completes a full cycle of SRG iteration. By repeating this procedure, one can finally obtain the value of partition function in the thermodynamic limit.

\subsection{Determination of the environment tensor $M^e$}

\begin{figure}
\includegraphics[width=0.43\textwidth]{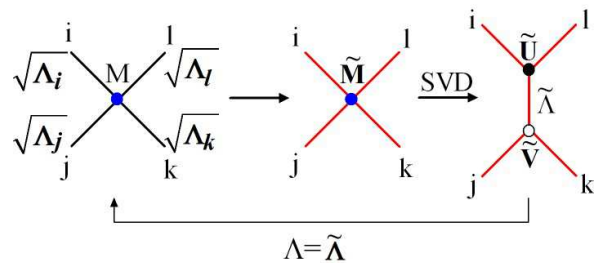}
\caption{Schematic representation of the iterative renormalization to the singular values $\Lambda$ of $M$. $\tilde{M}$ is obtained from Eq.~(\ref{eq:Me}) by taking a mean-field approximated $M^e$ defined by Eq.~(\ref{eq:MeanField}).}
\label{fig:srg-meanfield}
\end{figure}

In the DMRG calculation, a superblock can be separated into a system block and an environment block by cutting one bond. Thus the environment can be readily identified and integrated out. However, in the TRG method, it is highly nontrivial to handle the environment. To separate the environment from the system block, one needs to cut four bonds. The system contains only two sites and the environment contains all rest of sites. In this case, it is even more difficult to evaluate the contribution of the environment than the partition function itself in the TRG.

The key step in the SRG is to calculate the environment tensor $M^{e}$. In this respect, two approaches can be used. One is to take a mean-field approximation (or cavity approximation) to account for the environment contribution.\cite{poor-man} This is a cheap but less accurate approach. It is based on an intuitive interpretation to the singular values $\Lambda$ of $M$. The other is a more accurate approach. It is to evaluate $M^e$ directly from the environment lattice.

Let us first consider the ``mean field'' approach\cite{poor-man}. As mentioned before, the singular bond vector $\Lambda_n$ of $M$ defined by Eq.~(\ref{eq:Msvd}) is a measure of the entanglement between the corresponding basis states $U_n$ and $V_n$. It can be also regarded as a measure of the interaction between the two-end basis tensors $U$ and $V$ linked by this bond. If we assume that this singular vector also measures the entanglement between the system and the environment on all four dangling bonds $(i,j,k,l)$ connecting these two subsystems, then the environment matrix $M^e$ (up to an irrelevant prefactor) is approximately given by
\begin{equation}
\label{eq:MeanField}
M^e_{li,jk} \approx \sqrt{\Lambda_l \Lambda_i \Lambda_j \Lambda_k} .
\end{equation}
The reason we use $\Lambda^{1/2}$ instead of other exponent of $\Lambda$ as the weighing factor from the environment is because half of $\Lambda$ can be associated to the system and the other half to the environment. This is a simple but crude approximation. However, as will be shown later, it does reveal the importance of the renormalization effect of environment in the optimization of TRG.

\begin{figure}
\includegraphics[width=0.35\textwidth]{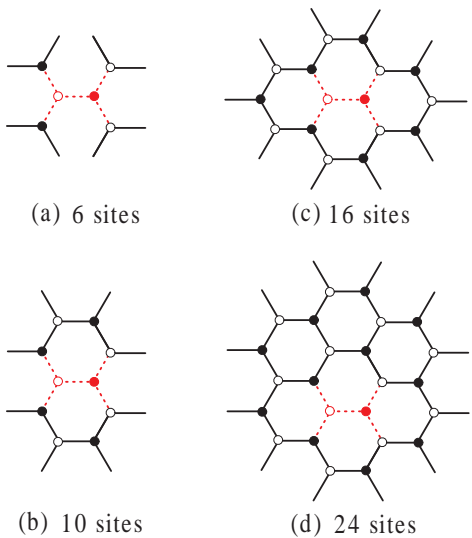}
\caption{Configurations of a finite honeycomb lattice with 6, 10, 16, or 24 sites. The corresponding environment lattice, by excluding the two red vertices, contains 4, 8, 14 and 22 sites, respectively. }
\label{fig:env-finite}
\end{figure}

To apply the above approximation, one needs to first calculate the singular vector $\Lambda$ of $M$, and then substitutes it into Eq.~(\ref{eq:MeanField}) to find $M^e$. From Eqs.~(\ref{eq:Me}) and (\ref{eq:Mtilde-svd}), one can find a new singular vector $\tilde{\Lambda}$. This $\tilde{\Lambda}$ incorporates partial contribution from the environment. To treat more accurately the environment contribution, one can replace $\Lambda$ in Eq.~(\ref{eq:MeanField}) with $\tilde{\Lambda}$ and repeat the above calculation iteratively. A graphic representation of this renormalization procedure is shown in Fig.~\ref{fig:srg-meanfield}.

In the above renormalization procedure, the truncation error of the partition function can be iteratively reduced. The iteration can be terminated when the truncation error is less than a desired value. In most of calculations, we find that two to three iterations are enough. However, at the critical point, more iterations are generally needed. The precision of the truncation error is not necessary to be set too high. If too many iterations are taken, some of the smallest singular values may become smaller than the machine error. In this case, further iterations will increase rather than reduce the error in the final result.

One can also treat self-consistently the above mean-field approximation, namely to require the bond vector $\Lambda$ used by $M^e$ in Eq.~(\ref{eq:MeanField}) to be the singular values defined in Eq.~(\ref{eq:Mtilde-svd}). This self-consistent approximation implies that the system is scaling invariant. It might be a good approximation at the critical point.

A more accurate and reliable approach is to evaluate $M^e$ from the environment lattice without taking the above mean-field approximation. For small lattice systems, for example, with only 6, 10, 16, and 24 sites as shown in Fig.~\ref{fig:env-finite} (the corresponding numbers of sites in the environments are 4, 8, 14, and 22, respectively), $M^e$ can be rigourously evaluated by contracting over all bond indices in the environment.

\begin{figure}
\includegraphics[width=0.45\textwidth]{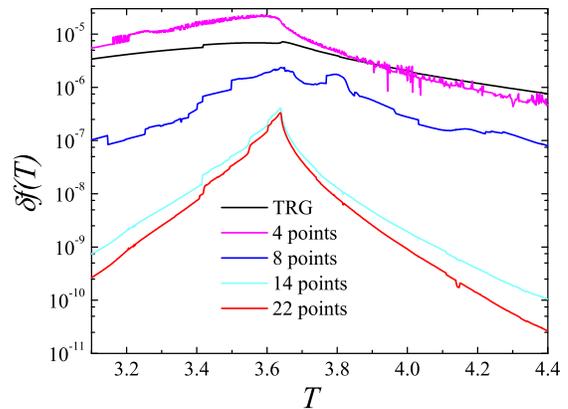}
\caption{Relative errors of the free energy for the Ising model on a triangular lattice obtained by considering the second renormalization effect from four finite environment lattices which contains 4, 8, 14, and 22 sites, respectively. The configurations of these environments are shown in Fig.~\ref{fig:env-finite}. The TRG result is also shown for comparison.}
\label{fig:error-finite}
\end{figure}

Fig.~\ref{fig:error-finite} shows how the relative error of the free energy,
\begin{equation}
\delta f(T) = 1 - \frac{f(T)}{f_{ex}(T)}
\end{equation}
varies with the environment size for the Ising model on a triangular lattice. $f(T)$ is the calculated free energy. $f_{ex}(T)$ is the exact free energy of the Ising model on a triangular lattice derived by Wannier\cite{Wannier}. The critical temperature of a paramagnetic to ferromagnetic phase transition is $T_c = 4/ \ln 3$. As expected, the relative error of the free energy decreases with the increase of environment lattice. Thus the renormalization effect of $M^e$ to $M$ can be more and more accurately described by increasing the environment size.

For a larger environment lattice, $M^e$ cannot be evaluated by directly tracing out all bond indices. An accurate and efficient approach we proposed in Ref.~\cite{Xie09} is to use the TRG to calculate $M^e$. This can be achieved by performing a forward iteration, followed by a backward iteration, described below:

\underline{Forward iteration:}

This is similar as in the standard TRG calculation. We apply the TRG to coarse grain the environment. At each step, one first constructs the M-matrix from Eq.~(\ref{eq:M}) by tracing out the common bond of two neighboring sites on which the $M$-matrix is defined. Then a singular value decomposition for this matrix is taken to find the corresponding $S^{a,n}$ and $S^{b,n}$ tensors using Eqs.~(\ref{eq:Sa}) and (\ref{eq:Sb}), where the superscript $n$ denotes the $n$'th TRG iteration.

\begin{figure}
\includegraphics[width=0.43\textwidth]{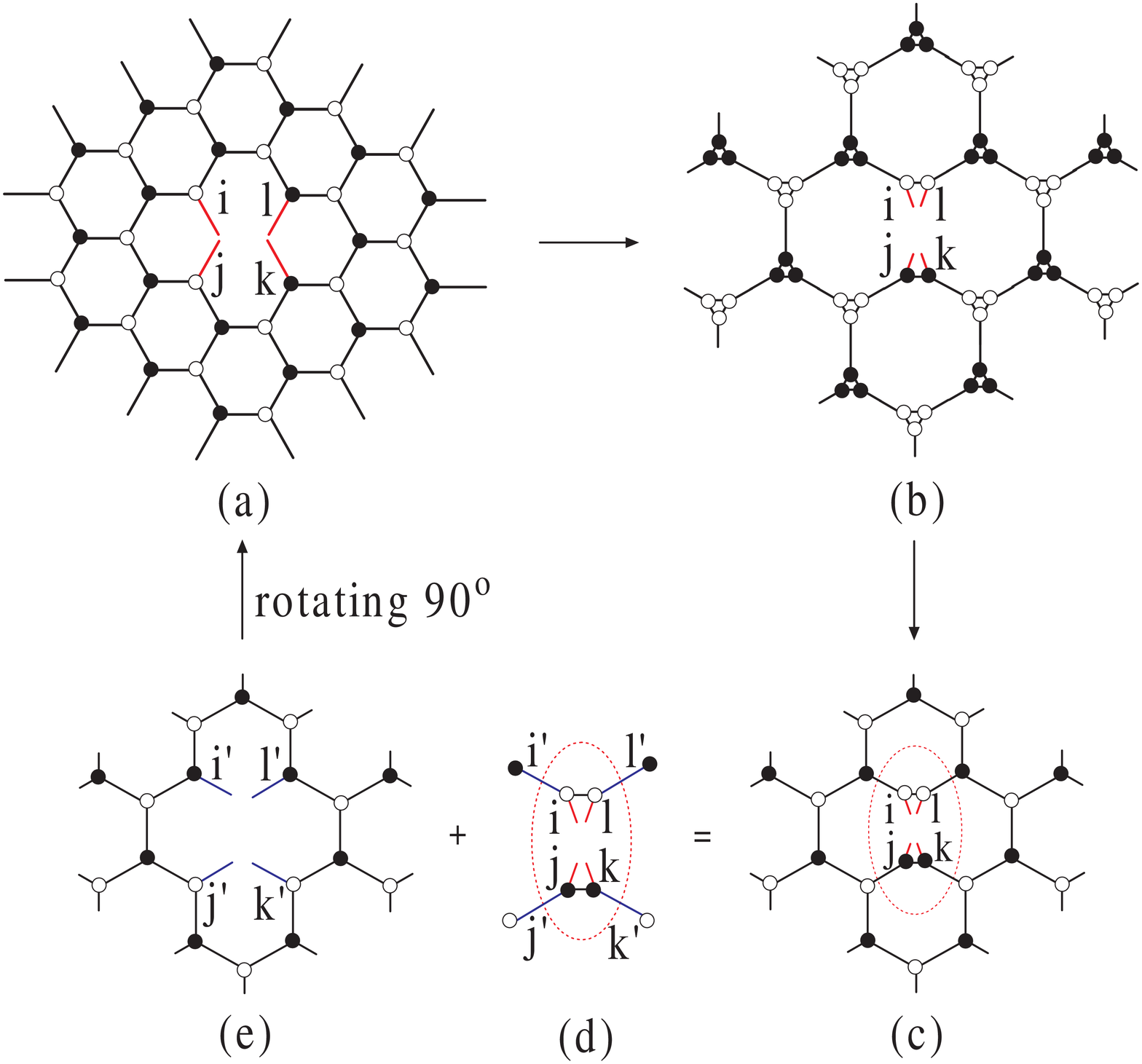}
\caption{One step of the forward iteration of the environment lattice.}
\label{fig:env}
\end{figure}

Fig.~\ref{fig:env} shows how the environment lattice changes at the $n$'th TRG iteration. For a given environment lattice as shown in Fig.~\ref{fig:env}(a), a rewired lattice whose configuration is given by Fig.~\ref{fig:env}(b) is derived by the singular value decomposition. A decimated lattice, shown in Fig.~\ref{fig:env}(c), is then obtained by contracting over the three internal bonds on each triangle and replacing it by a single lattice point. This new configuration of environment can be separated into two parts as
shown in Fig.~\ref{fig:env}(d) and (e). After rotating by 90 degrees clockwise, Fig.~\ref{fig:env}(e) looks exactly the same as the original environment lattice as shown in Fig.~\ref{fig:env}(a), except that the size of this reduced environment lattice is by a factor 3 smaller than its original one.

If we use $M^{(n-1)}$ to denote the environment matrix corresponding to the environment lattice shown in Fig.~\ref{fig:env}(a), then from Fig.~\ref{fig:env}, it is straightforward to show that it satisfies the following recursion formula
\begin{equation}
\label{eq:MeTRG}
M_{lijk}^{(n-1)}=\sum_{l^{\prime }i^{\prime }j^{\prime }k^{\prime
}}\sum_{pq}M_{l^{\prime }i^{\prime }j^{\prime }k^{\prime
}}^{(n)}S_{i^{\prime }pl}^{a,n}S_{j^{\prime }ip}^{a,n}S_{k^{\prime
}qj}^{b,n}S_{l^{\prime }kq}^{b,n} ,
\end{equation}
where $M^e= M^{(0)}$ is the final result to be found.

The above steps can be repeated until $M^e$ for a sufficiently large environment lattice can be accurately determined. If we assume that the above iteration is terminated at the $N$'th step so that only 4 sites are left in the reduced environment lattice as shown in Fig.~\ref{fig:env-finite}(a), then $M^{(N)}$ can be determined by the following formula
\begin{equation}
M_{lijk}^{(N)}=\sum_{abcd}T_{abl}^{a}T_{dic}^{b}T_{abj}^{b}T_{dkc}^{a}
\end{equation}
where the center-symmetric boundary conditions are assumed.
$T^{\alpha}$ ($\alpha = a, b$) is given by
\begin{equation}
T^\alpha_{xyz} = \sum_{ijk} S^{\alpha ,N}_{xji} S^{\alpha,N}_{ykj} S^{\alpha ,N}_{zik} .
\end{equation}

\underline{Backward iteration:}

In Eq.~(\ref{eq:MeTRG}), $M^{(n-1)}$ is determined by the environment matrix $M^{(n)}$ which is to be evaluated from the next iteration. Thus $M^{(n-1)}$ cannot be directly determined from the above forward iteration. However, we can do a backward iteration from $M^{(N)}$ to find the environment matrix $M^e = M^{(0))}$ using Eq.~(\ref{eq:MeTRG}).

From the above forward-backward iteration, one can calculate accurately the environment tensor $M^e$. In practical calculations, we find that there is no need to use a large $N$, since the environment tensor $M^{e}$ converges quickly with the increase of environment lattice.

\begin{figure}
\includegraphics[width=0.4\textwidth]{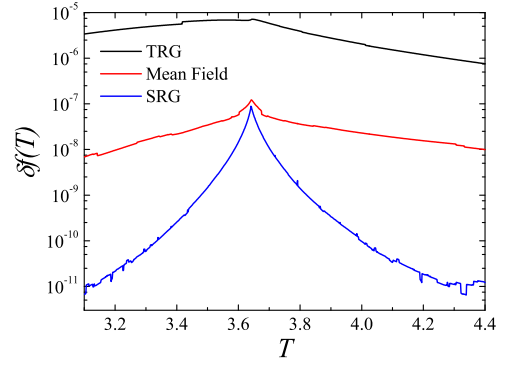}
\caption{(color online) Comparison of the relative error of the free energy for the Ising model on triangular lattices obtained using TRG (red), the mean-field approximated SRG (blue), and the SRG (black) methods with $D_{cut} = 24$, respectively. The critical temperature is $T_c = 4/ \ln 3$.}
\label{fig:error-triangle}
\end{figure}

Fig.~\ref{fig:error-triangle} compares the relative errors of the free energy for the Ising model on a triangular lattice obtained using the TRG, the mean-field approximated SRG, and the SRG methods, respectively. We find that the mean-field approach of SRG can improve dramatically the accuracy of the results. This shows that our intuitive assumption for the mean-field character of $\Lambda$ has indeed caught the main feature of the entanglement between the system and environment lattices. As expected, the improvement of SRG is even more impressive. It improves the accuracy for more than five orders of magnitude away from the critical point and for more than two orders of magnitude at the critical point by keeping only $D_{cut} = 24$ states. The accuracy can be further improved if the environment is determined self-consistently by SRG which, however, will increase the time cost.

\begin{figure}
\includegraphics[width=0.38\textwidth]{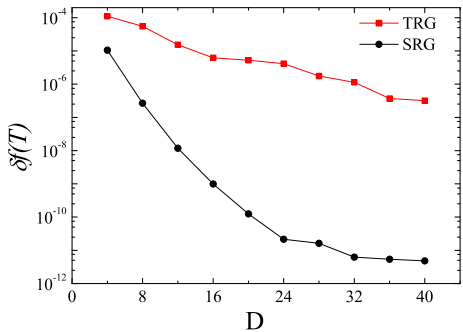}
\caption{(color online) The relative error of the free energy as a function of the truncation dimension $D_{cut}$ for the Ising model on triangular lattices obtained using the TRG (black) and SRG (blue),  respectively. $T = 3.2$}
\label{fig:error-vs-D}
\end{figure}

Furthermore, we find that the improvement of the SRG over the TRG becomes more and more pronounced with increasing $D$. Fig.~\ref{fig:error-vs-D} shows how the relative errors change with $D_{cut}$ for the Ising model on triangular lattices.

\subsection{Other lattices}

The above SRG approach can be extended to apply to other kinds of two dimensional lattices, such as square or Kagome lattices.

\begin{figure}[b]
\includegraphics[width=0.45\textwidth]{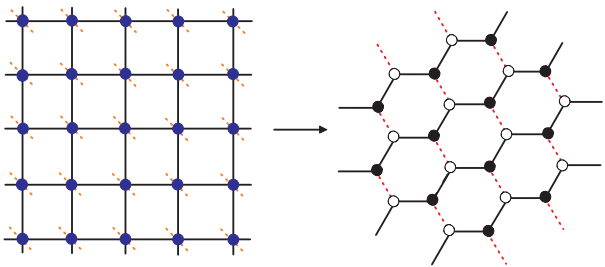}
\caption{To convert a square lattice (a) to a honeycomb lattice (b) by singular decomposing the fourth order tensor $T$ at each point into two separated third order tensors along the direction indicated by the red dashed lines. }
\label{fig:SqtoHoney}
\end{figure}

Let us first consider a square lattice. Unlike in a honeycomb lattice, the $M$-matrix is now the $T$-matrix defined at each lattice site. The TRG iterations on square lattices can be done using the approach introduced in Ref.~\cite{Levin07}. To evaluate the environment matrix $M^e$ at each TRG iteration, one can still use the forward-backward iteration scheme of SRG introduced on a honeycomb lattice. To do this, we need to first convert a square lattice to a honeycomb lattice. This can be done by singular decomposing the fourth order $T$-tensor defined  at each square lattice point into two separated third order tensors along a diagonal direction. A graphical representation of this decomposition of the square lattice is shown in Fig.~\ref{fig:SqtoHoney}.

After this conversion, the environment has the same configuration as
that for a honeycomb lattice shown in Fig.~\ref{fig:env}(a). The only difference is that in the present case the dashed bonds in Fig.~\ref{fig:SqtoHoney}(b) have higher dimensions than the solid bonds. Therefore, the forward-backward iteration scheme previously introduced can be used to calculate $M^e$ for a square lattice.

\begin{figure}[t]
\includegraphics[width=0.45\textwidth]{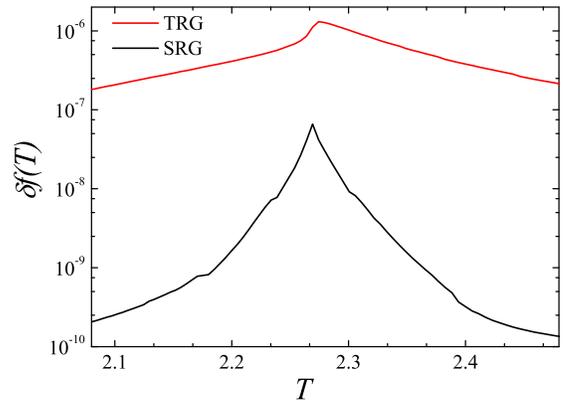}
\caption{Comparison of the relative errors of the free energy for the Ising model on a square lattice obtained by the SRG with those obtained by the TRG. $D=24$}
\label{fig:Sq-error}
\end{figure}

Fig.~\ref{fig:Sq-error} compares the relative errors of free energy for the Ising model on the square lattice obtained by the TRG with those obtained by the SRG. Similar as for the honeycomb lattice, we find that the SRG is much more accurate than the TRG. It improves the accuracy for more than three orders of magnitude away from the critical point and for more than one order of magnitude at the critical point over the TRG by keeping $D_{cut} = 24$ states.

\begin{figure}[b]
\includegraphics[width=0.45\textwidth]{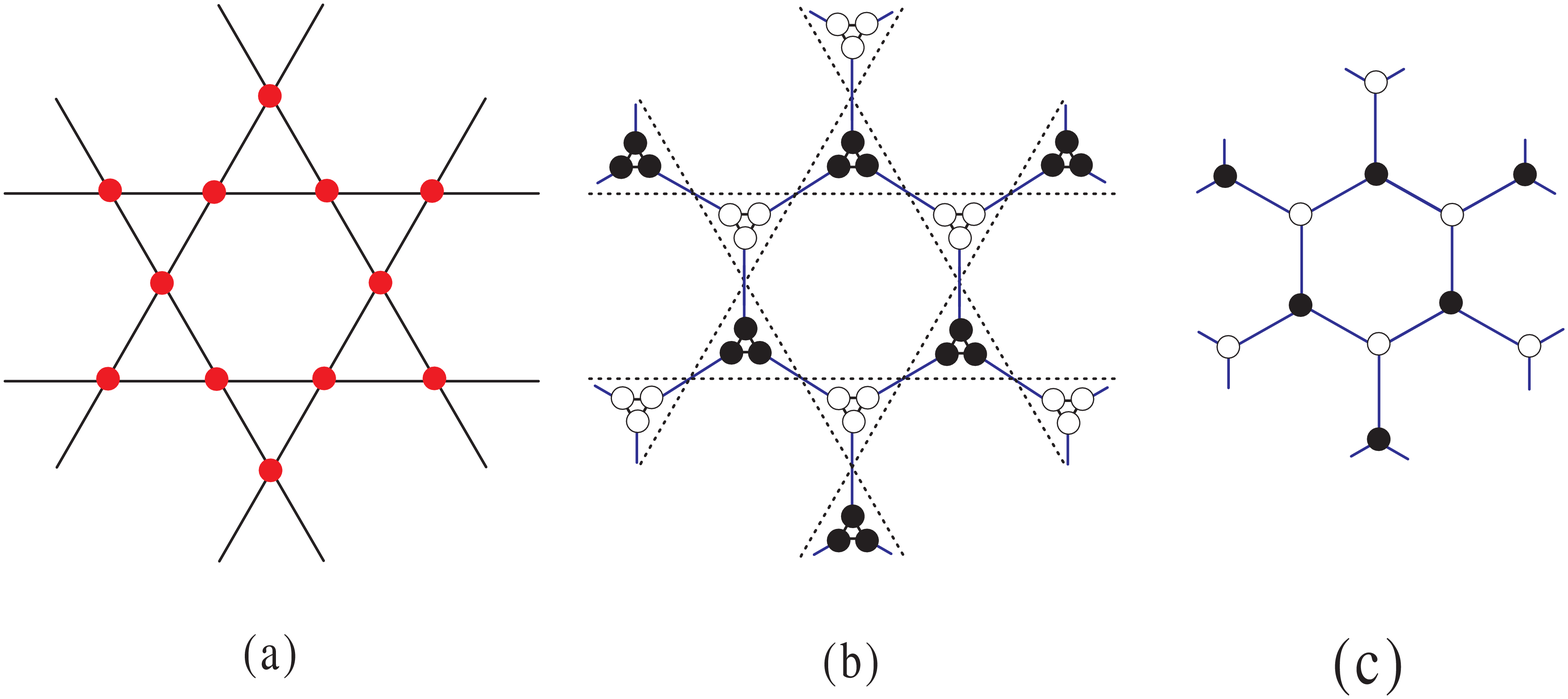}
\caption{ To convert a Kagome lattice (a) to a honeycomb lattice (c) by rewiring with singular value decompositions and contraction.}
\label{fig:Kagome}
\end{figure}

The SRG scheme introduced on the honeycomb lattice can be also applied to a tensor-network model defined on a Kagome lattice. A Kagome lattice can be converted to a hexagonal lattice in the following two steps: First, we singular decompose each local tensor to rewire the Kagome lattice (Fig.~\ref{fig:Kagome}a) to a triangular-hexagonal lattice (Fig.~\ref{fig:Kagome}b); Second, we contract all triangles in Fig.~\ref{fig:Kagome}(b) by tracing out all internal bonds on each triangle. This yields a honeycomb lattice as shown in Fig.~\ref{fig:Kagome}(c). Then one can calculate all thermodynamic quantities just using the method developed for the honeycomb lattice.

\section{Quantum Tensor Network States}
\label{sec:quantum}

As mentioned before, the wavefunction generated by the DMRG is a matrix product state. It can be shown that any wavefunction in one dimension can be faithfully expressed as a matrix product. In two or higher dimensions, a matrix product wavefunction is no longer a good description of the ground state, since the minimal matrix dimension that is needed for accurately describing the state grows exponentially with the lattice size.

A tensor network state is a natural extension of the one-dimensional matrix product state in two or higher dimensions. It captures the main feature of entanglement between different parts of wavefunction. For a quantum spin system, for example, a tensor network wavefunction can be expressed as:
\begin{equation}
|\psi \rangle = \text{Tr} \prod_{i} T^i_{x_i y_i z_i\cdots}[m_i] |m_i\rangle
\end{equation}
where $T^i_{x_i y_i z_i\cdots}[m_i]$ is the tensor defined at site $i$. The order of this tensor is the number of spins interacting with $S_i$. If there are only nearest-neighbor interactions, the rank of $T^i[m_i]$ is just the coordinate number of the lattice, which is 3 for a honeycomb lattice, 4 for a square or Kagome lattice, and 6 for a triangular lattice. The subscripts $(x_i, y_i, z_i \cdots)$ are bond indices. $m_i$ is a set of local basis states. The trace is to sum over all basis configurations and over all bond indices. The local tensor $T^i$ is generally site dependent. However, if the system is translational invariant, then $T^i$ can be site independent.

To determine the local tensor $T^i[m_i]$, a commonly adopted approach is to take all the tensor elements as variational parameters and to determine them by variationally minimizing the ground state energy
\begin{equation}
E = \frac{\langle \psi| H |\psi \rangle }{\langle \psi|\psi \rangle}.
\end{equation}
However, the number of variational parameters that can be efficiently handled by the minimization is only a few hundreds. This limits the bond dimension that can be handled to be generally less than 5 on a square lattice. Moreover, the accuracy of the wavefunction such obtained is very low.

In Ref.~\cite{Jiang08}, we proposed an iterative projection method to
determine the tensor-network wavefunction. This is a more efficient
and accurate method. Below we take the S=1/2 Heisenberg model on a
honeycomb lattice to demonstrate how the method works. The model
Hamiltonian is defined by
\begin{eqnarray}
H & = & \sum_{\langle ij \rangle} H_{ij}, \\
H_{ij} & = & J S_i \cdot S_j ,
\label{eq:model}
\end{eqnarray}
where $\langle ij \rangle$ stands for summation over nearest neighboring sites.

The honeycomb lattice is a bipartite lattice. It can be divided into two sublattices, denoted as $a$ and $b$, respectively. A translation invariant tensor-network wavefunction can in general be represented as:
\begin{equation}
\label{eq:wavefun}
|\psi \rangle = \text{Tr} \prod_{i\in a,j\in b}\lambda_{x_i} \lambda_{y_i} \lambda_{z_i} A_{x_i y_i z_i}[m_i] B_{x_j y_j z_j} [m_j]|m_i m_j\rangle
\end{equation}
where $A_{x_i y_i z_i}[m_i]$ and $B_{x_j y_j z_j}[m_j]$ are the tensors defined on the two sublattices, respectively. $\lambda_\alpha$ ($\alpha =x,y,z$) is a positive bond vector of dimension $D$, defined on the bond emitted from site $i$ along the $\alpha$ direction. $m_i$ is a local spin basis set of dimension $d$. The trace is to sum over all spin configurations and over all bond indices.

In Eq.~(\ref{eq:wavefun}), we introduce explicitly the bond vectors $\lambda$ in the tensor network wavefunction. These $\lambda$, as will be discussed below, measure approximately the entanglement between two neighboring sites. Since each bond connects with two tensors, one can regard each $\lambda$ as a product of two $\lambda^{1/2}$ and associate each of them to one of the tensors. In other words, one can take $\lambda^{1/2}$ as a mean field measure of entanglement acting from one tensor to another. In the calculation of the tensor network state by projection, the renormalization effect from the environment tensors need to be considered. This renormalization effect, similar as in the application of the SRG, can be taken into account approximately by these effective entanglement fields. This approximate treatment of the renormalization effect from the environment, as will be discussed below, works very well. It provides an accurate and efficient method for determining the tensor-network wavefunction.

In the projection method, the ground state wavefunction is determined by applying the projection operator $\exp (-\tau H)$ to an arbitrary initial state $|\Psi \rangle $ which is not orthogonal to the true ground state. In the limit $\tau \rightarrow \infty $, the resulting wavefunction will converge to the ground state of $H$. However, this projection cannot be done in a single step since the terms in $H$ defined by Eq.~(\ref{eq:model}) do not commute with each other. Instead, one needs to use a small $\tau $ and apply this projection operator to $|\Psi \rangle $ iteratively for sufficiently many times.

Let us start by dividing the Hamiltonian into three parts
\begin{eqnarray}
H &=&H_{x}+H_{y}+H_{z},
\nonumber \\
H_{\alpha } &=&\sum_{i\in a}H_{i,i+\alpha }\quad (\alpha =x,y,z).
\nonumber
\end{eqnarray}
$H_{\alpha }$ ($\alpha =x,y,z$) contains all the interaction terms along the $\alpha $-direction only. These terms commute with each other. For small $\tau$, one can use the Trotter-Suzuki formula to decouple approximately $\exp (-\tau H)$ into a product of three terms
\begin{equation}
e^{-\tau H} \approx e^{-\tau H_z} e^{-\tau H_y}e^{-\tau H_x} + o(\tau ^2) .
\end{equation}
A projection of $H$ can then be readily performed using $\exp (-\tau H_\alpha)$ ($\alpha = x, y, z$) in three steps.

In the first step, the projection is done with $H_x$. As $H_x$ contains only the interaction terms between two neighboring spins connected by horizontal bonds, this  projection generates a new wavefunction
\begin{eqnarray}
&& e^{-\tau H_x}|\Psi \rangle  \nonumber \\
&=& \mathrm{Tr} \prod_{i\in a, j=i+\hat{x}} \sum_{m_i m_j} \langle
m_i^\prime
m_j^\prime | e^{-H_{ij}\tau} |m_i m_j\rangle  \nonumber \\
&& \lambda_{x_i}\lambda_{y_i} \lambda_{z_i} A_{x_i y_i z_i} [m_i] B_{x_j y_j
z_j} [m_j] |m_i^\prime m_j^\prime \rangle,
\end{eqnarray}
where any two tensors linked by a horizontal bond are mixed by the matrix elements of the local projection operator. In order to perform the projection for the next step, one needs to separate these two tensors so that the wavefunction can return back to its original form.

To do this, let us define a $(D^2d)\times (D^2d) $ matrix using the two tensors on a horizontal bond and the bond vectors connected to these tensors
\begin{eqnarray}
&&S_{y_i z_i m_i^\prime , y_j z_j m_j^\prime}
\nonumber \\
& = & \sum_{m_i m_j}\sum_x \langle m_i^\prime m_j^\prime | e^{-H_{ij}\tau}| m_i m_j\rangle
\nonumber \\
&& \lambda_{y_i} \lambda_{z_i} A_{x y_i z_i}[m_i] \lambda_x B_{x y_j
z_j} [m_j] \lambda_{y_j} \lambda_{z_j} .
\label{eq:S1}
\end{eqnarray}
In this definition, the four bonds that connect to the environment, i.e. $y_i$, $z_i$, $y_j$ and $z_j$, are weighted by the corresponding bond vector $\lambda$, rather than $\lambda^{1/2}$. These extra $\lambda^{1/2}$'s are included on these four bonds to mimic the renormalization effect from the environment. This, similar as in our Poor Man's treatment of SRG, is a mean-field-type treatment to the environment. It is a key step in the application of this projection method. Without considering this renormalization effect, the iteration does not even converge.

We then take the singular value decomposition to decompose the above $S$ matrix into a product of two tensors at sites $i$ and $j$, respectively
\begin{equation}
S_{y_i z_i m_i , y_j z_j m_j} = \sum_x U_{y_i z_i m_i, x} \tilde{\lambda}_x V_{y_j z_j m_j,x}
\label{eq:svd}
\end{equation}
where $U$ and $V$ are two unitary matrices and $\tilde{\lambda}_x$ is a positive diagonal matrix of dimension $D^2d$. $\tilde{\lambda}_x$ is the singular value matrix of $S$. It measures the entanglement between $U$ and $V$ tensors.

To process the iteration, one has to truncate the basis space by keeping only the $D$ largest singular values of $\tilde{\lambda}_x$. The truncated $\tilde{\lambda}_x$ is then set as the new $\lambda_x$ for the next iteration. After this, tensors $A$ and $B$ are updated as follows
\begin{eqnarray}
A_{x y_i z_i} [m_i] & = & \lambda_{y_i}^{-1} \lambda_{z_i}^{-1} U_{y_i z_i m_i, x},  \label{eq:A} \\
B_{x y_j z_j} [m_j] & = & \lambda_{y_j}^{-1} \lambda_{z_j}^{-1} V_{y_j z_j m_j, x}.  \label{eq:B}
\end{eqnarray}

This completes the projection for all horizontal bonds. The projection for the bonds along the y- or z-direction can be done in the same way. By repeating this iteration procedure, an accurate ground state wave function can be finally determined. In practical calculation, we find that the projection converges more efficiently if the second-order Trotter-Suzuki decomposition formula is used to decouple the projection operators along different directions.

In Eq.~(\ref{eq:S1}), the renormalization effect of the environment is taken into account by a mean-field approximation. This approximate treatment does not minimize the truncation error at each step of projection. However, this does not affect the converging speed of the wavefunction, since the Trotter and truncation errors do not accumulate in the iteration of projection.

To see how well this projection method works, let us consider a one-dimensional system. In one dimension, one can do a canonical transformation to convert a matrix product wavefunction into a canonical form, in which the left and right subblocks are strictly orthogonal and $\lambda^2$ are the eigenvalues of the reduced density matrix\cite{canonical1D}. If this canonical transformation is applied to the matrix product state before each projection, then the truncation error is minimized and the wavefunction such obtained is the most accurate one.

\begin{figure}[tbp]
\includegraphics[width=0.45\textwidth]{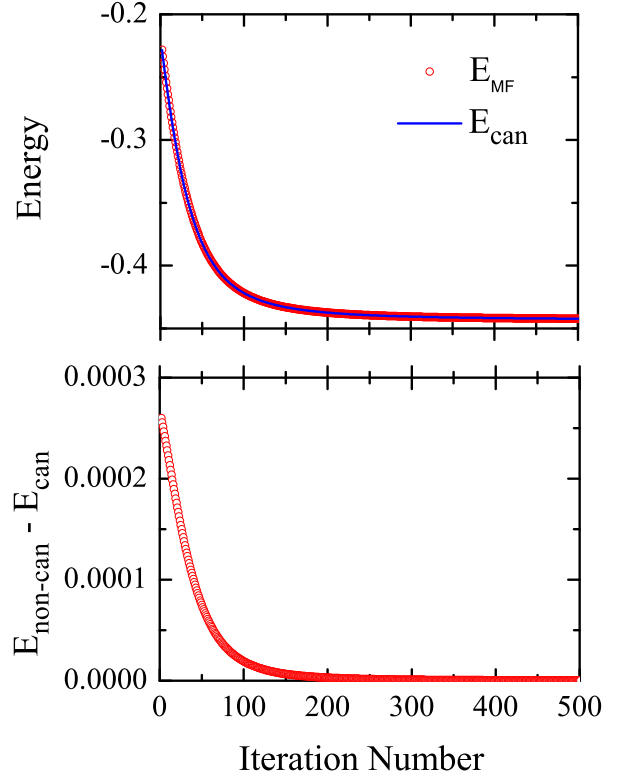}
\caption{(color online) Upper panel: Comparison of the ground state energy of the 1D Heisenberg model as a function of iteration number obtained using the projection method with and without taking canonical transformation for the matrix-product wavefunctions, $E_{can}$ and $E_{MF}$. Lower panel: the energy difference $E_{MF} - E_{can}$. The bond dimension $D=10$. }
\label{fig:1d}
\end{figure}

In the upper panel of Fig.~\ref{fig:1d}, we have compared the converging speed for the ground state energy obtained by taking the canonical transformation, $E_{can}$, with that obtained by simply using the mean-field approximation, $E_{MF}$, starting from a randomly generated wavefunction for the one-dimensional Heisenberg model with the bond dimension $D=10$. As can be seen from the figure, both $E_{can}$ and $E_{MF}$ converge to the true ground state energy of $D_{cut} = 10$ quickly. It is impossible to see the difference between $E_{can}$ and $E_{MF}$ on the energy scale of the figure. In the lower panel of the figure, we show the energy difference $E_{MF}-E_{can}$ as a function of the iteration number. At the beginning of the iteration, $E_{MF}$ is slightly higher than $E_{can}$, indicating that the wavefunction obtained by the canonical approach is indeed more accurate, as expected. However, the difference between these two energies approaches to zero with increasing number of iterations, indicating that the wavefunction obtained using the mean-field approach does converge to the true ground state wavefunction. The application of the mean-field method does not need to perform the canonical transformation at each step of projection. Thus in many cases it can be more efficient than the canonical approach in the evaluation of the wavefunction.

However, it should be pointed out that the above mean-field treatment to the environment cannot be simply applied to the calculation of time-dependent or thermodynamic quantities with the above projection method. This is because in the calculation of time-dependent or thermodynamic quantities, both the Trotter and truncation errors will accumulate with the step of iterations and the long-time (or imaginary time in the calculation of thermodynamic quantities) results may not converge. In this case, a more rigorous treatment to the renormalization effect from the environment should be considered to minimize the accumulated Trotter and truncation errors.

In practical calculation, the symmetry of the Hamiltonian can be used to block diagonalize tensors $A$ and $B$ in Eq.~(\ref{eq:wavefun}). This can improve greatly the efficiency of the calculation and allow larger bond dimension to be handled. For example, the conservation of the spin component along the $z$-axis, $S_z$, can be readily enforced by requiring the bond spins to satisfying the equation
\begin{equation}
S[x_i] + S[y_i] + S[z_i]  = m_i,
\label{eq:Sa}
\end{equation}
for the tensor on the $a$-sublattice, and the equation
\begin{equation}
-S[x_j] - S[y_j] - S[z_j]  = m_j.
\label{eq:Sb}
\end{equation}
for the tensor on the $b$-sublattice. In the above equations, $S[\alpha]$ is the spin quantum number of the bond index $\alpha$.

\begin{figure}[tbp]
\includegraphics[width=0.3\textwidth]{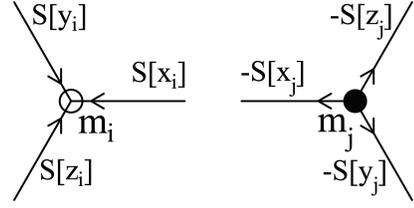}
\caption{(color online) Graphical representation of the spin conservation law defined by Eqs.~(\ref{eq:Sa}) and (\ref{eq:Sb}). The bond spin takes a positive/negative sign if the bond arrow points towards/outwords the vertex.}
\label{fig:symm}
\end{figure}

Graphically, the above spin conservation law can be represented by an arrow rule as shown in Fig.~\ref{fig:symm}. We put an arrow on each bond to represent the sign factor for the virtual bond spin. An inwards/outwards arrow takes a positive/negative sign. The conservation law is defined at each vertex, given by the equality that the sum of the bond spin times its arrow sign is equal to the physical spin at the vertex. This arrow rule can be extended to apply to all kinds of lattices, no matter they are bipartite or non-bipartite.

At each projection of a pair of tensors, for example the two tensors linked by a horizontal bond, $A_{x_i,y_i,z_i}$ and $B_{x_j,y_jz_j}$ ($x_j=x_i$), the conservation of total $S_z$ is satisfied if
\begin{equation}
S[y_i] + S[z_i] - m_i = S[y_j] + S[z_j] + m_j.
\end{equation}

From Eqs.~(\ref{eq:A}) and (\ref{eq:B}), it is simple to show that the sum of $S_z$ on the whole lattice is zero:
\begin{equation}
\sum_i m_i = 0 .
\end{equation}
Thus the tensor-network wavefunction satisfying the above conservation conditions has total $S_z = 0$. One can also construct a tensor-network wavefunction with nonzero total $S_z$ by modifying Eq.~(\ref{eq:A}) or (\ref{eq:B}) on some of the lattice points. But in this case, the wavefunction is no longer translation invariant.

It is straightforward to extend the above projection method to other quantum lattice models with short range interactions either in a honeycomb or other kinds of lattices in two or higher dimensions. At each step of projection, a singular value decomposition of a $D^{z -1}d \times D^{z -1}d$ matrix needs to be calculated, where $z$ is the coordinate number of the lattice. The largest bond dimension $D$ that can be studied depends on the coordinate number $z$. The larger $z$, the smaller $D$ that can be handled.

\section{Calculation of Physical Observables}
\label{sec:expect}

Now let us consider how to calculate the expectation values of a physical operator $\hat{O}$
\begin{equation}
\langle O \rangle  = \frac{\langle \Psi | \hat{O} |\Psi \rangle }{\langle \Psi |\Psi \rangle }.
\label{eq:exp}
\end{equation}
For a given tensor-network wavefunction $|\Psi\rangle$, it is simple to show that both the denominator and numerator can be expressed in the form of tensor products. In particular, the denominator has a similar form as for the partition function of a classical statistical model:
\begin{equation}
\langle \Psi | \Psi \rangle = \textrm{Tr} \prod_{i\in a,j\in b} T_{X_{i}Y_{i}Z_{i}}^{a} T_{X_{j}Y_{j}Z_{j}}^{b} ,
\label{eq:wf-tensor}
\end{equation}
where $T^a$ and $T^b$ are the local tensors on the two sublattices defined by
\begin{eqnarray}
T_{X_{i}Y_{i}Z_{i}}^{a} & = & \sum_{m}A_{x_{i}y_{i}z_{i}} [m] A_{x_{i}^{\prime} y_{i}^{\prime } z_{i}^{\prime }}^{\prime }[m] , \\
T_{X_{i}Y_{i}Z_{i}}^{b} & = & \sum_{m} B_{x_{i}y_{i}z_{i}}[m] B_{x_{i}^{\prime} y_{i}^{\prime } z_{i}^{\prime }}^{\prime }[m] ,
\end{eqnarray}
$X_{i} = (x_{i},x_{i}^{\prime })$, $Y_{i} = (y_{i},y_{i}^{\prime })$, and $Z_{i} = (z_{i},z_{i}^{\prime })$. The bond dimension of these local tensors is $D_c = D^2$.

\begin{figure}[tbp]
\includegraphics[width=0.45\textwidth]{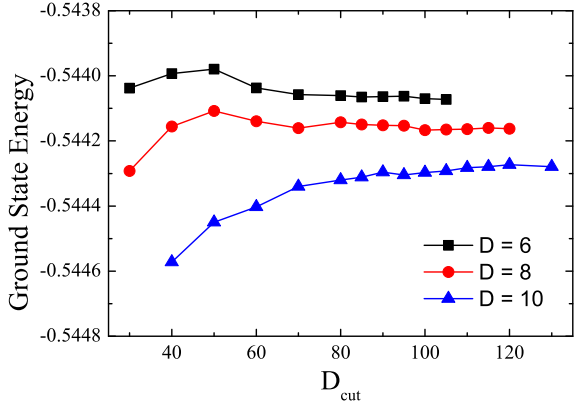}
\caption{(color online) The SRG result of the ground state energy as a function of the truncation dimension $D_{cut}$ for the Heisenberg model on a honeycomb lattice. $D$ is the bond dimension of the wavefunction.}
\label{fig:e_vs_Dcut}
\end{figure}

The numerator differs from the denominator only in the definition of the local tensors on the sites where the physical operators are defined. For example, in the evaluation of the ground state energy, one needs to calculate the expectation value of $H_{12}$:
\begin{equation}
\langle H_{1,2} \rangle = J\sum_{\alpha = x,y,z} \langle S_{1,\alpha} S_{2,\alpha} \rangle .
\end{equation}
This can be calculated from the sum of the above three terms on the right hand side. The corresponding local tensors on sites $1$ and $2$ are defined by
\begin{eqnarray}
T_{X_{i}Y_{i}Z_{i}}^{S_{1,\alpha}} & = & \sum_{m m^\prime } A_{x_i y_i z_i}[m] A_{x^\prime_{i}y^\prime_{i} z^\prime_{i}}[m^\prime ]
\langle m^\prime | S_{1,\alpha} |m\rangle , \nonumber \\
T_{X_{i}Y_{i}Z_{i}}^{S_{2,\alpha}} & = & \sum_{m m^\prime } B_{x_i y_i z_i}[m] B_{x^\prime_{i}y^\prime_{i} z^\prime_{i}}[m^\prime ]
\langle m^\prime | S_{2,\alpha} |m\rangle . \nonumber
\end{eqnarray}
The definition of local tensors on the other sites is unchanged. For the  Heisenberg model on the honeycomb lattice, the ground state energy per-site is given by $3 \langle H_{12} \rangle /2$.

The spin conservation law defined by Eqs.~(\ref{eq:Sa}) and (\ref{eq:Sb}) can be also implemented to tensors $T^a$ and $T^b$. If we define
\begin{equation}
S[X_i] = S[x_i] - S[x^\prime_i],
\end{equation}
and similarly for $S[Y_i]$ and $S[Z_i]$, then the spin conservation for $T^a$ or $T^b$ is then given by
\begin{equation}
S[X_i] + S[Y_i] + S[Z_i] = 0 .
\end{equation}

The expectation value (\ref{eq:exp}) can be calculated using the SRG method\cite{Xie09, Jiang08} introduced in Sec. \ref{sec:SRG}. Fig.~\ref{fig:e_vs_Dcut} shows how the ground state energy varies with the truncation dimension $D_{cut}$ for the Heisenberg model on a honeycomb lattice. For the three cases shown in the figure, the ground state energy shows a small variance at the fifth decimal when $D_{cut}$ is above 80. This variance results from the truncation error in the SRG calculation. The ground state energy does not decrease monotonically with $D_{cut}$, because SRG is not a variational approach.

\begin{figure}[tbp]
\includegraphics[width=0.42\textwidth]{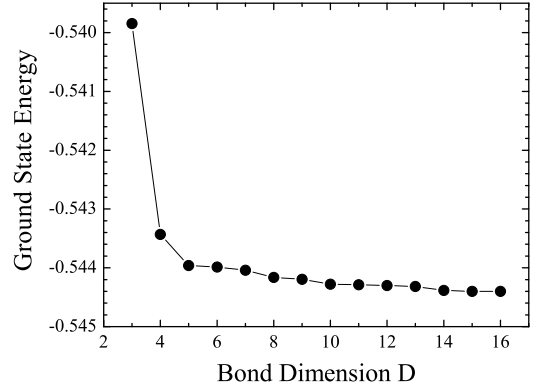}
\caption{(color online) The ground state energy of the Heisenberg model on a honeycomb lattice as a function of the bond dimension $D$ obtained by the SRG with $D_{cut}=130$.}
\label{fig:ge}
\end{figure}

Fig.~\ref{fig:ge} shows the ground state energy of the Heisenberg model as a function of the bond dimension $D$ obtained using the SRG. The result converges quickly with increasing $D$. This is because the ground state energy is determined by the local correlation function. It is insensitive to the long wavelength fluctuation of the wavefunction. The converged ground state energy of $D=16$ is -0.54440 by keeping $D_{cut}=130$ states. It agrees with the most recent Monte Carlo result $E = -0.54455 (20) $.\cite{Low09} It is also consistent with the spin wave\cite{Zheng91} (-0.5489) as well as the series expansion\cite{Otimaa92} (-0.5443) results.

For the Hamiltonian, or any other physical operators which commute with $H$, the ground state $|\Psi\rangle$ can be a common eigenfunction of these variables. The expectation values of these conserving physical variables can be also evaluated by the following equation
\begin{equation}
\langle O \rangle  = \frac{\langle \Phi | \hat{O} |\Psi \rangle }{\langle \Phi |\Psi \rangle },
\end{equation}
where $|\Phi\rangle$ is an arbitrary wavefunction that is not orthogonal to $|\Psi\rangle$. An advantage for evaluating the expectation value using this formula is that the bond dimension of $|\Phi\rangle$ can be much smaller than $|\Psi\rangle$. This can reduce the bond dimension $D_c$ for the tensors used in Eq.~(\ref{eq:wf-tensor}) and allow a tensor-network wavefunction with a relatively larger bond dimension to be studied.

\begin{figure}[tbp]
\includegraphics[width=0.42\textwidth]{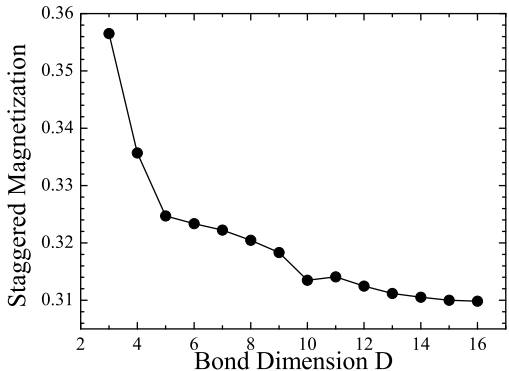}
\caption{(color online) The staggered magnetization as a function of $D$ for the Heisenberg model on a honeycomb lattice.}
\label{fig:Mstag}
\end{figure}

For the Heisenberg model on a honeycomb lattice, the ground state is spontaneous symmetry broken. It possesses a long range antiferromagnetic order with a finite staggered magnetization. The staggered magnetization measures the long range spin-spin correlation functions. In an applied staggered magnetic field, the spin wave excitation is gapped. The spin-spin correlation function, excluding a constant long range term, decays exponentially with their distance. However, in the absence of an applied staggered magnetic field, the spin wave excitation is a gapless Goldstone mode. The low energy, or long wavelength, spin fluctuation is strong in a Honeycomb lattice. These low energy fluctuations can affect strongly the behavior of the staggered magnetization in the low field limit.

In the ground state of which the SU(2) spin rotation symmetry is broken by applying a staggered magnetic field  $h_s$ (assuming along the $z$-axis), the staggered magnetization can be evaluated from the formula
\begin{equation}
M_{s} = \frac{1}{2} \langle S_{1,z} - S_{2,z} \rangle .
\end{equation}

Fig.~\ref{fig:Mstag} shows the SRG results\cite{note} for the staggered magnetization $M_s$ as a function of the bond dimension $D$ in the zero field limit $h_s \rightarrow 0$. In contrast to the ground state energy, $M_s$ converges slowly with increasing $D$. When $D = 16$, we find that $M_s \approx 0.3098$, which is higher than the recent quantum Monte Carlo result\cite{Low09}, $M_s \approx 0.2681(8)$. It is also higher than the result obtained from the spin wave theory\cite{Zheng91}, $M_s = 0.24$, or the series expansion\cite{Otimaa92}, $M_s = 0.27$. This difference is due to the quantum fluctuation in the ground state. In the ground state, the spin excitation is gapless and the spin-spin correlation is long ranged. However, for a tensor-network wavefunction, this long-range spin-spin correlation is terminated by the finite bond dimension. This is a drawback of the tensor product wavefunction in the studying of a gapless state. This kind of error can be reduced by increasing the bond dimension of the tensor product state or by adopting other kinds of tensor-network wavefunctions.

\begin{figure}[tbp]
\includegraphics[width=0.42\textwidth]{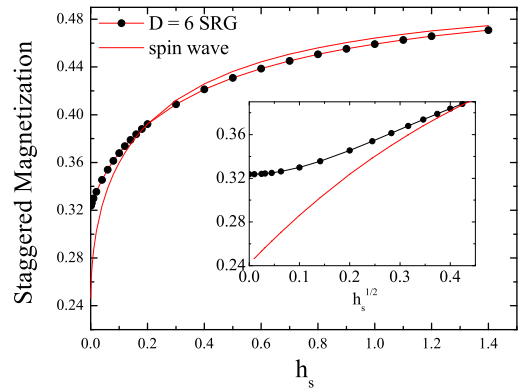}
\caption{(color online) The field dependence of the staggered magnetization $M_s$ for the Heisenberg model on a honeycomb lattice. The inset shows the $h_s^{1/2}$ dependence of $M_s$. }
\label{fig:Mstag}
\end{figure}

To understand this more clearly, we compare the SRG result for the staggered magnetization as a function of an applied staggered field $h_s$ with that obtained from the spin-wave theory in Fig.~\ref{fig:Mstag}. At high field, the two results agree qualitatively with each other. This is because in a finite field, the spin wave excitation is gapped and the entanglement entropy satisfies the area law. In this case, the tensor-network wavefunction is a good approximation to the true ground state. However, in the limit $h_s\rightarrow 0$, the spin wave excitation becomes gapless and the critical spin fluctuation becomes important. In particular, as shown in the inset of Fig.~\ref{fig:Mstag}, the staggered magnetization obtained by the spin wave theory varies as $\sqrt{h_s}$ in the low field limit. This $\sqrt{h_s}$ dependence of the staggered magnetization is due to the low-energy (or long wavelength) spin excitations. The long wavelength correlation is not included in the tensor-network approximation of the ground state. It leads to the error in the SRG result of the zero-field staggered magnetization for finite $D$.

\begin{figure}[tbp]
\includegraphics[width=0.5\textwidth]{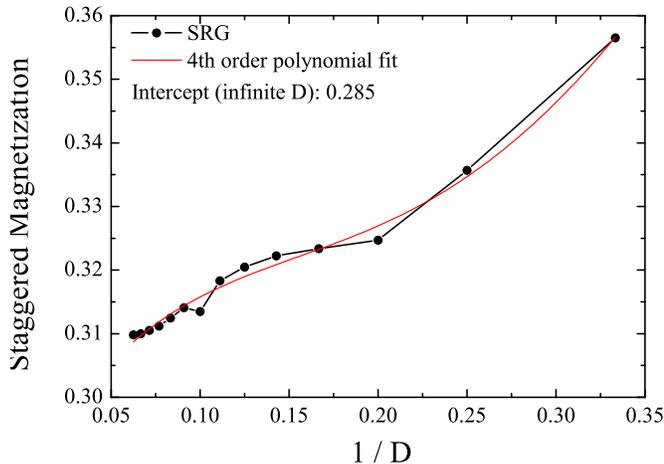}
\caption{(color online) The staggered magnetization as a function of $1/D$ for the Heisenberg model on a honeycomb lattice.}
\label{fig:MsinverseD}
\end{figure}

The staggered magnetization in the zero field limit can nevertheless be more accurately estimated by extrapolating the SRG result to the limit $D\rightarrow \infty$. Fig.~\ref{fig:MsinverseD} shows the fourth order polynomial fit to the SRG results of staggered magnetization as a function of $1/D$. The extrapolated staggered magnetization in the limit $1/D \rightarrow 0$ is about 0.285, in agreement qualitatively with the quantum Monte Carlo result, $M_s \approx 0.2681(8)$, as well as the series expansion one, $M_s = 0.27$.

\section{Summary}
\label{sec:summary}

In this paper, we have discussed the tensor-network representation of  classical statistical models, and given a comprehensive introduction to the iterative projection and the SRG methods. A classical statistical model with local interactions can be represented as a tensor-network model either in its original lattice or in its dual lattice. The order of local tensor is equal to the coordinate number of the lattice on which the tensor-network model is defined. In practical calculation, one should choose a tensor-network representation in which the order of local tensors is the smallest. For example, in the study of a classical model in the honeycomb/triangular lattice, the tensor-network model defined in the original/dual lattice should be used.

The projection method introduced in Sec.~\ref{sec:quantum} is an efficient and accurate tool for evaluating the tensor-network wavefunction for a quantum lattice model. It allows a tensor-network state with a large bond dimension, for example $D=70$ in a honeycomb lattice, to be determined. In the projection, the renormalization effect of the environment is taken into account by a mean-field treatment of the bond vector. This reduces significantly the truncation error at each step of projection and enable the wavefunction to converge fast with the increase of iterations. As both the Trotter and truncation errors do not accumulate in the iteration of projection, the accuracy of the wavefunction can be well controlled by adjusting the Trotter parameter $\tau$.

The SRG is an accurate numerical method for evaluating thermodynamic properties of classical tensor-network models as well as the expectation values of quantum tensor-network states. It generalizes the TRG method of Levin and Nave\cite{Levin07} to account for the renormalization effect from the environment in the decomposition of local tensors. This method reduces dramatically the truncation error and improves significantly the accuracy of TRG. The concept of SRG is ubiquitous. The key idea is to maximize the entanglement between the system and environment in the basis truncation. It can be applied to the tensor network models. It can be also extended to apply to other physical problems where the system can be divided into two parts and the interplay between them is important.

Both the projection method and the SRG can be applied to a finite lattice system. They can also be applied to a system without any translation or rotation symmetry. In this case, one has to treat each tensor independently. The cost (both the CPU time and memory space) scales linearly with the lattice size.

The tensor network wavefunction provides a good description for the ground states of two-dimensional quantum lattice models. For the spin-1/2 Heisenberg model on a honeycomb lattice, the ground state energy obtained from the tensor network states converges fast with the bond dimension $D$. Our SRG results for $D \le 16$ already reach the accuracy of the recent quantum Monte Carlo calculation. By further increasing $D$ or reducing the truncation error in the SRG calculation, we believe that more accurate results for the ground state energy will be obtained in near future.

The tensor network state satisfies the entanglement area law. It captures the key feature of short range correlations. The correlation function between any two local operators in a tensor network state is always short ranged. In other words, all low energy excitations are gapped in a tensor network state. However, long range correlations are not correctly described by a tensor network state with finite bond dimension. This leads to a relatively large error in the determination of physical quantities, such as the staggered magnetization in the honeycomb Heisenberg model, whose values are governed by gapless low-lying excitations. To resolve this problem, one has either to increase the bond dimension or to adopt a new type of tensor network wavefunction, such as the multiscale entangled tensor network state\cite{Vidal-mera}.

The iterative projection and SRG methods introduced in this paper improve significantly the accuracy and efficiency in the study of tensor-network states/models. These methods can be used to investigate the ground state properties of interacting fermions\cite{Banuls} or frustrated quantum spin models. It also has the potential to be extended to probe thermodynamic as well as dynamic properties of quantum lattice models in two dimensions. The application of these methods is still in its early stage.\cite{Chen09,Li10} Further development of these methods may enlighten the route for the exploration of new numerical renormalization group methods, and lead to the solution of a number of problems that are difficult to be solved by other methods.

\section*{Acknowledgement}

We thank H.C. Jiang, W. Li, B. Normand, and Z.Y. Weng for helpful discussions. This work was supported by the NSF-China and the National Program for Basic Research of MOST, China.

\end{document}